\documentclass[aps,prl,12pt]{revtex4}

\usepackage{graphicx}

\begin{document}

\title{The Rayleigh--Taylor instability and internal waves in quantum plasmas}

\author{Vitaly Bychkov, Mattias Marklund and Mikhail Modestov }

\affiliation{Department of Physics, Ume\aa\ University, SE--901 87
Ume\aa, Sweden}

\begin{abstract}
Influence of quantum effects on the internal waves and the
Rayleigh-Taylor instability in plasma is investigated. It is shown
that  quantum pressure always stabilizes the RT instability. The
problem is solved both in the limit of short-wavelength
perturbations and exactly for density profiles with layers of
exponential stratification. In the case of stable stratification,
quantum pressure modifies the dispersion relation of the inertial
waves. Because of the quantum effects, the internal waves may
propagate in the transverse direction, which was impossible in the
classical case. A specific form of pure quantum internal waves is
obtained, which do not require any external gravitational field.
\end{abstract}

\maketitle

Studies of quantum plasmas was initiated by Pines in the 1960's \cite{Pines,Pines-book},
where the finite width of the electron wave function gives rise to dispersion, important in
the  high-density and/or low temperature regime.
A number of quantum plasma studies has since appeared Ref.\ \cite{kremp-etal}, \textit{e.g.},
kinetic models of the quantum electrodynamical properties of nonthermal
plasmas \cite{bezzerides-dubois} and covariant Wigner function descriptions of relativistic
quantum plasmas \cite{hakim-heyvaerts}. There has recently been a surge in the
interest of quantum plasmas, see \textit{e.g.}
Refs.\ \cite
{Manfredi2005,haas-etal1,haas,shukla,garcia-etal,marklund-brodin,brodin-marklund,BM-pairplasma},
in particular the nonlinear properties of dense \cite{Shukla-Eliasson1,Shukla-Eliasson2,Shaikh-Shukla} or magnetized plasmas \cite{BM-pairplasma,brodin-marklund-pre,marklund-etal}. Many of these studies have motivated by new discoveries concerning
nanostructured materials \cite{craighead} and quantum wells
\cite{manfredi-hervieux}, the discovery of ultracold plasmas
\cite{robinson-etal,li-etal,fletcher-etal},
astrophysical applications \cite{harding-lai}, and inertial fusion plasmas \cite{glenzer-etal}.
For such quantum systems, the so called
Bohm--de Broglie potential \cite
{Manfredi2005,haas-etal1,haas,shukla,garcia-etal}, as well as the zero
temperature Fermi pressure
\cite{Manfredi2005,haas-etal1,haas,shukla,garcia-etal} and other spin
properties \cite{marklund-brodin,brodin-marklund,BM-pairplasma,brodin-marklund-pre,marklund-etal}
can significantly modify the dynamics of the plasma. Moreover,
quantum electrodynamical effects can give rise to completely new effects in plasma environments \cite
{marklund-shukla,Lundin2007,lundstrom-etal,Brodin-etal-2007}
that can be of relevance in high-intensity quantum plasmas.
Within the fluid approach to quantum plasmas  \cite
{Manfredi2005,haas-etal1,haas,shukla,garcia-etal,marklund-brodin,brodin-marklund,Lundin2007,lundstrom-etal,Brodin-etal-2007}, collective effects can be described within a
unified picture.

The above examples mainly focuses on oscillations 
in homogeneous quantum plasma backgrounds. However, quantum plasmas can in practice often involve nonuniform
density profiles, which often develop in an real (\textit{e.g.} in astrophysics) 
or effective (\textit{e.g.} in inertial
confined fusion) external gravitational
field. In the classical case, stratified plasma in a
gravitational field inevitably exhibits either inertial waves or
the Rayleigh-Taylor (RT) instability depending on whether the
stratification is stable or unstable \cite{Landau-Lifshitz-Fluid}.
The purpose of the present paper is to study influence of the
quantum effects on the internal waves and the RT instability. Here
we show that quantum pressure always stabilizes the RT
instability. The stabilization has a meaning of an effective
"quantum velocity" reducing the instability growth rate. In that
sense the stabilization is similar to the RT stabilization by an
ablation flow in inertial confined fusion
\cite{Bychkov.et.al-1991,Bychkov.et.al-1994,Betti.et.al-1996,Sanz.et.al-2006}.
We solve the problem in the limit of short-wavelength
perturbations. We also find exact solutions to the RT stability
problem for density profiles with layers of exponential
stratification. In the case of stable stratification, quantum
pressure modifies the dispersion relation of the inertial waves.
Because of the quantum effects, the internal waves may propagate
in the transverse direction, which was impossible in the classical
case. Even more, we obtain specific form of pure quantum internal
waves, which do not require any external gravitational field. The results
could be of significance for astrophysical and intertial fusion plasmas.

We start with equations for continuity and momentum transport in
quantum plasmas in the MHD approximation taking into account
gravitational field
\cite{Gardner-1994,Haas2005,Marklund-Brodin2007,Brodin-Marklund2007}
\begin{equation}\label{eq:cont}
  \frac{\partial \rho}{\partial t} + \nabla\cdot(\rho\mathbf{u}) = 0 ,
\end{equation}
\begin{eqnarray}
  && \rho\left( \frac{\partial}{\partial t} + \mathbf{u}\cdot\nabla
  \right)u_{l}
    =-\frac{\partial P}{\partial x_{l}} + \rho g_{l}+
  \nonumber \\ &&\qquad\qquad
   \frac{\hbar^2}{12m_em_i}
\frac{\partial}{\partial x_{j}}
    \left(\rho\frac {\partial^{2} }{\partial x_j\partial x_l}\ln{\rho}\right) ,
  \label{eq:mom}
\end{eqnarray}
where $m_{e}$, $m_{i}$ are electron and ion masses. We consider
small-scale effects taking the gravitational acceleration to be
$\mathbf{g} = - g\hat{\mathbf{z}}$, where $g$ is a constant. The
hydrodynamic equilibrium is determined by the
 balance of forces
\begin{equation}
  \frac{dP_0}{dz} = -\rho_0g + \frac{\hbar^2}{12m_em_i}\frac{d}{dz}\left(
    \rho_0 \frac{d^2}{d z^2}\ln{\rho_0}
  \right)
\end{equation}
However, the hydrodynamic equilibrium does not necessarily imply
thermodynamic equilibrium. In the incompressible limit of an
essentially subsonic plasma dynamics it allows both stable and
unstable density profiles.

We consider small perturbations of the equilibrium according to
$\varphi(x,z,t) = \varphi_0(z) + \tilde{\varphi}(z)\exp(i\omega t
+ ikx)$, where $\varphi$ denotes any of the fluid variables, $k$
is the perturbation wave number, and $\omega$ is the wave
frequency. In the case of unstable stratification the frequency
should be replaced by the instability growth rate $\sigma = i
\omega$. Here we are interested in the incompressible plasma
dynamics with
\begin{equation}\label{eq:incompressible}
  \frac{d\rho}{dt} = 0
\end{equation}
for any Lagrangian plasma parcel.  Incompressible flow is typical
both for the RT instability and the internal waves
\cite{Kull-1991};  equation (\ref{eq:incompressible}) holds for
$\omega/k \ll c_s$, where $c_s = [(\partial
P/\partial\rho)_S]^{1/2}$  is the sound speed. Then, to first
order in the perturbed quantities, the system
(\ref{eq:cont})--(\ref{eq:mom}) reads
\begin{equation}\label{eq:cont1}
  i\omega\tilde{\rho} + \tilde{u}_z\frac{d\rho_0}{dz} = 0 ,
\end{equation}
\begin{equation}\label{eq:cont2}
  \frac{d\tilde{u}_z}{dz} + ik\tilde{u}_x = 0 ,
\end{equation}
\begin{eqnarray}
  && i\omega \rho_0\tilde{u}_x = -ik\tilde{P}+
  \nonumber \\ &&\qquad
   i k \frac{\hbar^2}{12m_em_i}\left\{\frac{d}{dz}\left[\rho_{0}\frac{d}{dz}\left(\frac{\tilde{\rho}}{\rho_{0}}\right)
  \right] -k^{2} \tilde{\rho} \right\},
  \label{eq:mom1}
\end{eqnarray}
and
\begin{eqnarray}
  && i\omega \rho_0\tilde{u}_z = -\frac{d\tilde{P}}{dz} - \tilde{\rho}g
  +
  \nonumber \\ && \qquad
    \frac{\hbar^2}{12m_em_i}\frac{d}{dz}\left[\rho_{0}\frac{d^{2}}
  {dz^{2}}\left(\frac{\tilde{\rho}}{\rho_{0}}\right)+\tilde{\rho}\frac{d^{2}}
  {dz^{2}}\ln \rho_{0} \right]-
  \nonumber \\ && \qquad\qquad
    k^{2}\frac{\hbar^2}{12m_em_i}\rho_{0}\frac{d}{dz}\left(\frac{\tilde{\rho}}{\rho_{0}}\right).
  \label{eq:mom2}
\end{eqnarray}

We start with the limit of short wavelength perturbations,
 $k/\alpha \gg 1$, where $\alpha = d\ln \rho_0/dz$
is the local inverse length scale of the density profile. In that
case we can use the Wentzel-Kramers-Brillouin method
\cite{Landau-Lifshitz-Quantum} presenting perturbations in the
form $\varphi(x,z,t) = \varphi_0(z) +\tilde{\varphi}\exp(ik_xx +
ik_zz + i\omega t)$. Then, in the short wavelength limit  Eqs.\
(\ref{eq:cont2}) - (\ref{eq:mom2}) reduce to
\begin{equation}\label{eq:cont2-short}
  ik_{z}\tilde{u}_x + ik_{x}\tilde{u}_x = 0 ,
\end{equation}
\begin{equation}
  i\omega \rho_0\tilde{u}_x = -ik_{x}\tilde{P}-
i k_{x} k^{2}\frac{\hbar^2}{12m_em_i}\tilde{\rho},
  \label{eq:mom1-short}
\end{equation}
\begin{eqnarray}
  && i\omega \rho_0\tilde{u}_z = -ik_{z}\tilde{P} - \tilde{\rho}g
  +
  \nonumber \\ && \qquad
  k^{2}\frac{\hbar^2}{12m_em_i}\tilde{\rho}
  \left(\frac{1}{\rho_{0}}\frac{d\rho_{0}}{dz}-ik_{z}\right).
  \label{eq:mom2-short}
\end{eqnarray}
Solving equations (\ref{eq:cont1}), (\ref{eq:cont2-short}) -
(\ref{eq:mom2-short}) we find the dispersion relation for the
internal waves in quantum plasma
\begin{equation}\label{eq:dispersion}
  \omega^{2}=\omega_{cl}^{2}\frac{k_{x}^{2}}{k^{2}}+
  k_{x}^{2}\frac{\hbar^2}{12m_em_i}\left(\frac{1}{\rho_{0}}\frac{d\rho_{0}}{dz}\right)^{2},
\end{equation}
where the classical frequency of internal waves is designated by
$\omega_{cl}^{2} = -(g/\rho_0)d\rho_0/dz > 0$. In the case of
"horizontal" internal waves, $k_z = 0$, $k_{x}=k$, the dispersion
relation takes the form similar to electrostatic and
electromagnetic plasma waves
\begin{equation} \label{eq:disp-horiz}
  \omega^{2} =
    \omega_{cl}^2 + U_{q}^2 k^{2},
\end{equation}
where $\omega_{\text{cl}}$ plays the same role as the plasma
frequency and
\begin{equation}\label{eq:quant-speed}
  U_q = \frac{\hbar}{2\sqrt{3m_em_i}}
      \left|\frac{1}{\rho_0}\frac{d\rho_0}{dz}\right|
\end{equation}
is the characteristic quantum speed. Even in the case of zero
gravity with $\omega_{cl}=0$ we obtain specific quantum internal
waves with the dispersion relation
\begin{equation}\label{eq:disp-quant}
  \omega = U_qk_x.
\end{equation}

In the case of unstable stratification $\mathbf{g}\cdot\nabla\rho
< 0$, or
\begin{equation}\label{eq:factor}
  \frac{g}{\rho_0}\frac{d\rho_0}{dz} > 0 ,
\end{equation}
 we obtain the RT instability instead of internal waves.
The RT perturbations of short wavelength are strongly localized
within the layer with the most steep density profile $ \max \alpha
= d\ln\rho_0/dz$, e.g. see
\cite{Bychkov.et.al-1991,Bychkov.et.al-1994}, and the maximal
instability growth rate corresponds to the mode with $k_{z}=0$
\begin{equation} \label{eq:RT-horiz}
  \sigma =
    \sqrt{g\alpha - U_{q}^{2}k^{2}}.
\end{equation}
As we can see from (\ref{eq:RT-horiz}), quantum effects always
play a stabilizing role for the RT instability. In a sense, this
stabilization is similar to the effects of ablation flow and
thermal conduction in the context of the inertial confined fusion
\cite{Bychkov.et.al-1991,Bychkov.et.al-1994,Betti.et.al-1996,Sanz.et.al-2006}.
Still, the ablation flow may be destabilizing for perturbations of
long wavelength $k/\alpha \ll 1$ because of the additional
Darrieus-Landau instability of a deflagration front
\cite{Bychkov.et.al-1994,Sanz.et.al-2006}. On the contrary,
quantum effects are always stabilizing.

We can also find exact analytical solutions to Eqs.\
(\ref{eq:cont1}) - (\ref{eq:mom2}) for certain density profiles.
For example, exact solution is possible for plasma consisting of
one or several layers with exponential stratification
$\rho_{0}\propto\exp(\alpha z)$, $d(\ln\rho_{0})/dz=\alpha=const$.
After tedious but straightforward algebra, the system Eqs.\
(\ref{eq:cont1}) - (\ref{eq:mom2}) may be reduced to a single
equation
\begin{equation} \label{eq:RT-exp}
  \frac{d}{dz}\left(\rho_{0}\frac{d\tilde{u}_{z}}{dz}\right)
  +\left(\frac{g}{\sigma_{\text{eff}}^{2}}\frac{d\rho_{0}}{dz}-\rho_{0}\right)k^{2}\tilde{u}_{z}=0,
\end{equation}
which has the same mathematical form as the respective classical
result \cite{Kull-1991} but with the "effective" growth rate
modified by the quantum terms
\begin{equation} \label{eq:sigma-eff}
\sigma_{\text{eff}}^{2}=\sigma^{2} + U_{q}^{2}k^{2}.
\end{equation}
Therefore, reproducing calculations for the RT instability growth
rate in the  classical case with  $\sigma_{cl}$, we find the
respective result for quantum plasma as
\begin{equation} \label{eq:sigma-q}
\sigma^{2}=\sigma_{cl}^{2} - U_{q}^{2}k^{2}.
\end{equation}
Below we give some  examples for the solution to Eq.
(\ref{eq:RT-exp}).

\paragraph*{1) \textit{Bounded plasma layer.}}

We consider density profile $\rho_{0}\propto\exp(\alpha z)$ within
the layer $0<z<L$  bounded by two rigid walls. The boundary
conditions at the walls are  $\tilde{u}_{z}=0$. Then we have
$\tilde{u}_{z}\propto \exp(\mu z)$ and Eq. (\ref{eq:RT-exp})
reduces to
\begin{equation} \label{eq:RT-mu}
\mu^{2}\tilde{u}_{z}+\alpha \mu \tilde{u}_{z}
\left(\frac{g\alpha}{\sigma_{\text{eff}}^{2}}-1\right)k^{2}\tilde{u}_{z}=0
\end{equation}
with the solutions
\begin{equation} \label{eq:RT-mu-1}
\mu_{1,2}= - \frac{\alpha}{2}\pm
\left(\frac{\alpha^{2}}{4}+k^{2}-\frac{g\alpha}{\sigma_{\text{eff}}^{2}}k^{2}\right)^{1/2}
\end{equation}
 The boundary condition $\tilde{u}_{z}=0$  at  $z=0, L$ is satisfied for
\begin{equation} \label{eq:RT-bc}
\left(\frac{g\alpha}{\sigma_{\text{eff}}^{2}}k^{2}-\frac{\alpha^{2}}{4}-k^{2}\right)^{1/2}=\frac{\pi
n}{L},
\end{equation}
which leads to the instability growth rate
\begin{equation} \label{eq:RT-sigma}
\sigma^{2}=g\alpha
\left(1+\frac{\alpha^{2}}{4k^{2}}+\frac{\pi^{2}n^{2}}{k^{2}L^{2}}\right)^{-1}-U_{q}^{2}k^{2}.
\end{equation}
In the limit of short wavelength perturbations we recover our
previous result Eq. (\ref{eq:RT-horiz}); the mode number $n$ has
the physical meaning of the scaled wave number in the vertical
direction, $k_{z}= \pi n /L$.

\paragraph*{2) \textit{A transitional layer.}}

\begin{figure}
\includegraphics[width=.9\columnwidth]{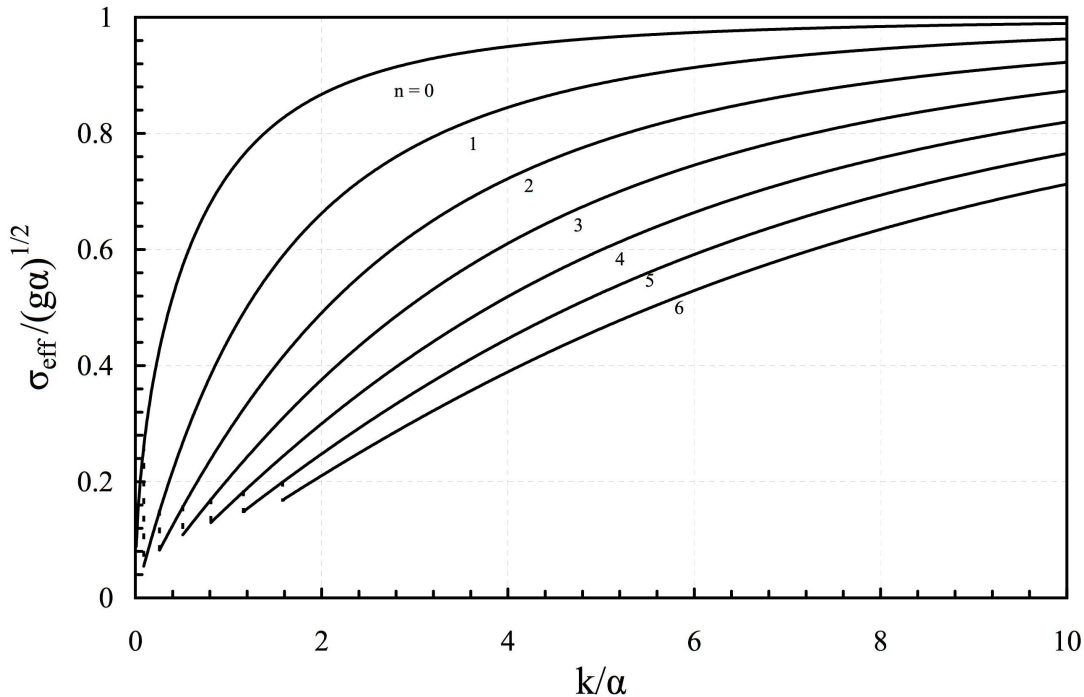}
\caption{Solution to Eq. (\ref{eq:trans}),
$\sigma_{\text{eff}}/\sqrt{g\alpha}$, versus the scaled wave
number $k/\alpha$ for the density drop $\Theta = 8$. Different
branches of the plot at $k>k_{c}$ correspond to different
mode numbers $n=0-6$.} \label{fig-1}
\end{figure}

\begin{figure}
\includegraphics[width=.9\columnwidth]{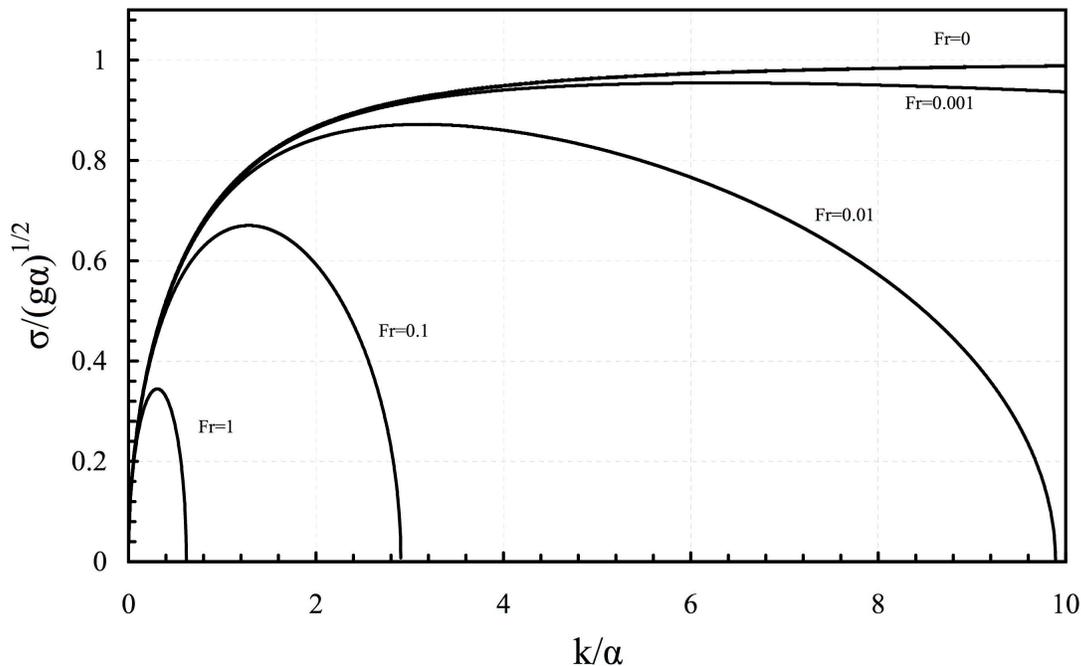}
\caption{Scaled instability growth rate $\sigma/\sqrt{g\alpha}$,
Eq. (\ref{eq:trans}), versus the scaled wave number $k/\alpha$ for
the density drop $\Theta = 8$  and the quantum Froude number
$\text{Fr}_{q}= 0-1$.} \label{fig-2}
\end{figure}

We consider a transitional layer of width $L$ separating two uniform plasmas of
different density $\rho_{1}$ and $\Theta \rho_{1}= \rho_{1} \exp(\alpha L)$
\begin{eqnarray}
  &&\rho=\rho_{1}\qquad\qquad\qquad\text{for}\qquad z<0,
  \nonumber \\
  &&\rho=\rho_{1}\exp(\alpha z)\qquad\text{for}\qquad 0<z<L,
  \nonumber \\
  &&\rho=\Theta\rho_{1}\qquad\qquad\qquad\text{for}\qquad z>L.
  \label{eq:layer}
\end{eqnarray}
In the uniform layers we obtain $\tilde{u}_{z}\propto \exp(\pm k
z)$ from Eq. (\ref{eq:RT-exp}), which leads to the boundary
conditions
\begin{equation} \label{eq:bound}
\frac{d\tilde{u}_{z}}{dz}=\pm k \tilde{u}_{z}
\end{equation}
at   $z=0, L$, respectively. Solving Eq. (\ref{eq:RT-exp}) with
boundary conditions Eq. (\ref{eq:bound}) we come to a
transcendental equation
\begin{equation} \label{eq:trans}
\frac{g\alpha
}{2\sigma_{\text{eff}}^{2}}=1+\frac{\beta}{k}\coth(\beta L),
\end{equation}
where
\begin{equation} \label{eq:sigma-trans}
\beta=
\left(\frac{\alpha^{2}}{4}+k^{2}-\frac{g\alpha}{\sigma_{\text{eff}}^{2}}k^{2}\right)^{1/2}.
\end{equation}
Solution to Eq. (\ref{eq:trans}) is unique and $\beta$ is real for
\begin{equation} \label{eq:critical k}
k<k_{c}=\frac{\alpha}{2}\left(\sqrt{1+4/\ln^{2}\Theta}-2/\ln\Theta\right).
\end{equation}
For $k>k_{c}$ the value $\beta$ becomes
imaginary, and we obtain multiple solutions to Eq.
(\ref{eq:trans}) similar to (\ref{eq:RT-sigma}), see Fig. \ref{fig-1}.
Different branches of the plot at $k>k_{c}$ correspond to
different mode numbers $n=0-6$, which show the number of
zeros for the eigenfunction $\tilde{u}_{z}$. In the limit of short
wavelength perturbations, $k/\alpha \gg 1$, $kL \gg 1$,
 Eq. (\ref{eq:trans}) goes over to (\ref{eq:RT-sigma}). In the opposite limit of
long wavelength perturbations,  $k/\alpha \ll 1$, $kL \ll 1$,   we
find
\begin{equation} \label{eq:sigma-long}
\sigma^{2}=\frac{\Theta - 1}{\Theta + 1}gk - U_{q}^{2}k^{2}.
\end{equation}
The first term in Eq. (\ref{eq:sigma-long}) is the RT instability
growth rate at the discontinuity, while the second one stands for
quantum stabilization.  In principle, the term with quantum
stabilization also contains small parameter  $k/\alpha \ll 1$, and
it should be omitted within the model of discontinuous density.
However, the quantum term contains also another dimensionless
parameter
\begin{equation} \label{eq:Froude-q}
\text{Fr}_{q}=\frac{U_{q}^{2}\alpha}{g}=\frac{\hbar^2\alpha^{3}}{12m_em_i
g},
\end{equation}
which plays the role of a quantum Froude number, and which may be
either large or small depending on a particular problem. For this
reason, the quantum term in Eq. (\ref{eq:sigma-long}) may be
important even in the limit of long wavelength perturbations.
Numerical solution to  Eq. (\ref{eq:trans}) is shown in Fig.
\ref{fig-2} for different values of the quantum Froude number. In
the domain $k>k_{c}$ the solution is not unique; in Fig.
\ref{fig-2} we presented the mode providing  the
maximal growth rate. As we can see from Fig.
\ref{fig-2}, in the case of the quantum Froude number above unity, $Fr_{q}>1$, 
stabilization happens already for perturbations of short wavelengths, $k/\alpha <1$.

%========================================================================%
%                              Figures                                   %
%========================================================================%
%\newpage
%========================================================================%
%========================================================================%
%========================================================================%
%========================================================================%

The effect of quantum dispersion on the RT instability
and internal waves in inhomogeneous systems
can be of relevance in astrophysical environments \cite{harding-lai} and
ultracold plasmas \cite{robinson-etal}. The stabilizing effect due
to the collective version  of Heisenberg's uncertaintly relation
could even dominate the dynamics in very dense plasmas. However,
for direct applications to such systems, more detailed calculations have to be made.
Here we have indicated the principle dynamics of inhomogeneous
fluids where such quantum effects can play a major role.
Generalizations and more detailed calculations, incorporating \textit{e.g.} magnetic pressure, are left for future research.

This work has been supported in part by the Swedish Research
Council (VR) and by the Kempe Foundation.

\end{document}